# Generation of Multipeak Spectrum of Spin Torque Oscillator in Non-linear Regime

Shuichi Iwakiri,[1] Satoshi Sugimoto,[2] Yasuhiro Niimi,[1,3] Kensuke Kobayashi,[1,4] Yusuke Kozuka,[2] Yukiko K. Takahashi[2], and Shinya Kasai[2,5]

[1]*Department of Physics, Graduate School of Science, Osaka University, 1-1 Machikaneyamacho, Osaka 560-0043, Japan*

[2]*Research Center for Magnetic and Spintronic Materials, National Institute for Materials Science (NIMS), 1-2-1 Sengen, Tsukuba 305-0047, Japan*

[3]*Center for Spintronics Research Network (CSRN), Graduate School of Engineering Science, Osaka University, 1-3 Machikaneyamacho, Osaka 560-8531, Japan.*

[4]*Institute for Physics of Intelligence and Department of Physics, The University of Tokyo, 7-3-1 Hongo, Tokyo 113-0033, Japan*

[5]*JST, PRESTO, 4-1-8 Honcho, Kawaguchi, Saitama 332-0012, Japan*

## Abstract

**We investigate the spectral characteristics of spin torque oscillator (STO) excited by the spin Hall-induced spin current. We observe that the modest spin current injection triggers the conventional single peak oscillating behavior of STO. As the spin current is further increased to enter the non-linear regime, we find the transition of the spectrum from a single peak to multipeak structure whose frequency spacing is constant. This behavior can be primarily explained by the extremely broadened peak of the STO, which is accompanied by the frequency-dependent filtering by the transmission line. To explain the observation more quantitatively, we also discuss that the multipeak may reflect the characteristics of**

**the intrinsic dynamics of STO in the non-linear regime.**

Nanoscale spintronic devices [1] can be versatilely manipulated by the spin current, which is generated by charge current via spin polarization in ferromagnets [2], the spin Hall effect in nonmagnetic heavy metals [3,4,5], and so on. The spin current exerts a torque (spin torque) to magnetization in adjacent ferromagnets, which can excite wide variety of magnetization dynamics such as magnetization switching and auto-oscillation (spin torque oscillator, STO) [6,7,8,9,10,11,12,13,14,15,16,17].

STO shows the characteristic spectrum in the non-linear regime [15]. One of the significant examples is that it exhibits equally-spaced multiple peaks when modulated externally by a feedback circuit, AC signal injection, etc. [18,19,20,21]. Such a spectrum stems from temporal modulation of the magnetization dynamics by means of magnetic field or AC spin-torque provided by the feedback. STO with such characteristic spectrum can potentially find its applications in frequency-comb, communication devices, or neuromorphic computation [22,23,24]. Therefore, it would be preferable if the multipeak spectrum can be realized without active additional circuit, which decreases complexity and energy consumption of STO-based devices.

In this Letter, we study the spectral behavior of an STO as a function of spin Hall-induced spin-torque. Besides a conventional single peak oscillation, we have observed equally-spaced multipeak at large spin torque regime without the help of any active circuit. The overall behavior of the spectrum can be explained by the extreme peak broadening in non-linear regime, which is accompanied by the frequency filtering effect due to the transmission line. We also point out that an additional mechanism such as the intrinsic non-linear dynamics of STO may also play a role in generating the multipeak spectrum.

Figure 1(a) shows a schematic of the experimental setup. A multilayered stack consisting of Ru (3.0 nm) / Ta (5.0 nm) / $Co_{20}Fe_{60}B_{20}$ (4.0 nm) / MgO (1.0 nm) / $Co_{20}Fe_{60}B_{20}$ (1.3 nm) / Ta (7.0 nm) was deposited from top to bottom on a thermally oxidized Si substrate using magnetron sputtering. The number in (…) defines the thickness of each layer and $Co_{20}Fe_{60}B_{20}$ is denoted as CoFeB from now on. The film was then annealed at 300 °C for 30 min. The CoFeB (4.0 nm) / MgO (1.0 nm) / CoFeB (1.3 nm) layer forms a magnetic tunnel junction (MTJ) with in-plane magnetic anisotropy as shown Fig. 1(a). The film stack above the bottom Ta layer (7.0 nm) was patterned into an elliptical element with 80 nm in a long radius and 40 nm in a short radius. The bottom Ta layer (7.0 nm) is used as a spin current source through the spin Hall effect, which exerts a torque on the magnetization of the bottom CoFeB (1.3 nm) layer [8,9]. Figure 1 (b) shows the magnetoresistance of the MTJ measured at room temperature. The magnetoresistance ratio and resistance area product are typically 75% and 30 $\Omega\mu m^2$, respectively. Figure 1 (c) shows the frequency characteristics of the transmission line solely, that is, the $S_{11}$ of the line measured by a vector network analyzer (VNA). As shown here, $S_{11}$ periodically increase and decrease as a function of the frequency, which we will discuss later to interpret our experimental result.

The direct current (DC) through the bottom Ta layer ($-0.5$ mA $\leq I_{Ta} \leq 1.5$ mA) excites the auto-oscillation of the magnetization through the spin Hall effect. A small current through the MTJ ($I_{MTJ}$ = 0.1 mA) was also applied from the DC port of the bias-tee to obtain the oscillation spectrum relevant to the magnetization dynamics. The in-plane magnetic field of 20 mT was applied by 60 degrees tilted from the long axis of the pillar (see Fig .1(a)). All the measurements were performed at room temperature.

We first explain the overall characteristics of the spectrum. Figures 2(a), 2(b), 2(c), and 2(d) show the power spectral density (PSD) measured at $I_{MTJ}$ = 0.1 mA for $I_{Ta}$ = 0.0, 0.5, 1.0, and, 1.5 mA, respectively. The background signals were removed by subtracting the spectrum obtained at $I_{MTJ} = I_{Ta} = 0$ mA. In the low positive current regime (0 mA ≤ $I_{Ta}$ ≤ 0.5 mA), the peak located at 680 MHz (main peak) grows, as shown in Figs. 2(a) and 2(b). As shown in Fig. 2 (c), with further increasing $I_{Ta}$ to 1.0 mA, the peak shifts to the lower frequency (redshift) and becomes broader (peak broadening). Such features are commonly observed in STO [25,26].

When $I_{Ta}$ is further increased up to 1.5 mA, a spectrum consisting of several peaks (multipeak spectrum) appears as shown in Fig. 2(d). We call this region the nonlinear regime, as the spin current is much larger than the threshold to trigger the conventional oscillation. Figure 2(e) shows the color plot of the PSD amplitude as functions of the frequency and $I_{Ta}$. The peak grows only for positive $I_{Ta}$, which guarantees that the auto-oscillation is excited by the spin torque. The frequency spacing of the peaks is estimated to be ~ 105 MHz. Note that such a multipeak spectrum has been known so far only in the STO modulated by active external circuit [20,21]. The observation of the multipeak spectrum free from active components is the central experimental finding in this Letter.

From now on, we discuss the mechanism of the multipeak spectrum observed in this regime. One clue is that $S_{11}$ of the transmission line has periodic structure in every 105 MHz as shown in Fig.1 (c), which corresponds to the frequency spacing in the multipeak spectrum. This suggests that the coaxial cable transmission line connecting between the MTJ and the amplifier (Fig. 1 (a)) works as a microwave cavity. Especially, the impedance mismatch at each end (50 Ω for the transmission line and ~3,000 Ω for the MTJ) makes the transmission of the signal sensitively limited to particular frequencies corresponding to the cable length. In addition to this fact, it is important to note that the peak broadening of STO is possibly to be enhanced more than 10 times in this non-linear regime [27]. In our case, such extreme peak broadening enables a peak to cover as much as a few hundreds of MHz, which involves several frequencies that $S_{11}$ takes its local maximum/minimum. Therefore, the origin of the multipeak generation is likely to be attributed to this extreme peak broadening in non-linear regime in combination with the filtering effect by the transmission line.

According to this scenario, we estimate the total power and the linewidth of the spectrum, to characterize the broadened peak. Figure 3 (a) shows the total power obtained by numerically integrating the whole spectrum. Corresponding to the spectral change shown in Fig.2 (e), the total power behaves differently depending on the value of $I_{Ta}$. For $I_{Ta} \leq 0.5$ mA, it increases monotonically as $I_{Ta}$ increases. In this region, the single mode around 680 MHz is mainly exited by the spin torque. Right after this region, a kink appears at $I_{Ta} \sim 0.5$ mA as denoted by an arrow in Fig.3 (a). In the following region ($I_{Ta} > 0.5$ mA), the STO gradually enters the non-linear regime, *i.e.* the spectrum starts to show the redshift and the broadening. There is also a slight kink at $I_{Ta} \sim 1.2$ mA, which corresponds to the multipeak generation as explained below.

Similar behavior can be caught in the linewidth of the spectrum (full width at half maximum; FWHM) shown in Fig. 3 (b). For $I_{Ta} \leq 0.5$ mA, FWHM is obtained by fitting a peak around 570 and 680 MHz by a single Lorentzian (center frequency, linewidth, and peak power are set as free parameters). For $I_{Ta} > 0.5$ mA, FWHM is estimated from the envelope of the spectrum (as an example, see the dotted line in Fig.2 (d)). As shown in Fig.3 (b), the linewidth is almost constant (~30 MHz) for $I_{Ta} \leq 0.5$ mA, while it starts to increase gradually for $I_{Ta} > 0.5$ mA. The linewidth has a kink at $I_{Ta} = 1.2$ mA, where the PSD of the individual peaks in multipeak structure exceeds 2 pW/MHz, making each peak well separable. This analysis captures the characteristics of the extreme peak broadening in non-linear regime, supporting its relevance to the origin of the multipeak in combination with the filtering effect by the transmission line.

The above scenario, however, requires additional mechanism to more quantitatively explain the observed phenomena. According to the above, the ratio of the apparent peak and dip (local maximum/minimum of the spectrum) value seen in the multipeak spectrum should correspond to $S_{11}$ of the transmission line. As the ratio of this peak/dip value of $S_{11}$ (defined as voltage ratio of the incident and reflection signal) is $\frac{10^{-1.05}}{10^{-1.15}} \sim 1.2$ (ex. 470 MHz for peak and 575 MHz for dip), the power ratio of the peak/dip should be no more than $\sim (1.2)^2 = 1.44$. On the other hand, the ratio calculated from the oscillation power spectrum (Fig.2 (d)) is larger than 3. This quantitative discrepancy indicates that the above scenario is not fully sufficient to explain our observation.

We point out that the intrinsic non-linear dynamics of STO is necessary to solve this discrepancy. Recently, we calculate the STO dynamics in the numerical simulation using the Landau-Lifshitz-Gilbert equation [28], where the similar values of the parameters (magnitude and anisotropy of the magnetization) in our experiment are adopted. We just focus on the dynamics of STO itself without a transmission line. We find that, in the non-linear regime, the magnetization of STO behaves in a complicated manner beyond a simple single mode oscillation and as a result the spectrum indeed transits from single peak to multipeak structure which is equally spaced (frequency spacing ~ 100 MHz). Although the range of the spin current for this simulation is not exactly the same as in our experiment, such intrinsic mechanism is feasible to be responsible for our experimental findings.

In conclusion, we have observed the multipeak spectrum of STO in non-linear regime without the help of any external active circuits. We show that the extreme peak broadening with frequency-dependent filtering by the transmission line is responsible for the observed multi-peak spectrum. The possible relevance of the intrinsic non-linear dynamics is also pointed out. This work extends the controllability of STO dynamics with spin torque, enabling us to realize non-linear oscillator-based applications in a simple manner.

**Acknowledgments**

This work is partially supported by JSPS KAKENHI Grant Nos. JP17K18892, JP18J20527, JP19H00656, JP19H05826, JP16H05964, and JP26103002, RIEC, Tohoku

University. The authors acknowledge Y. Suzuki and T. Taniguchi for fruitful discussions. The authors also acknowledge M. Takahagi for technical support.

**Data Availability**

The data that support the findings of this study are available from the corresponding author upon reasonable request.

**Figure Captions**

**Figure 1:** (a) Schematic of the experimental setup. MTJ nanopillar consists of CoFeB/MgO/CoFeB grown on the Ta underlayer. The spin Hall effect in Ta converts the charge current $I_{Ta}$ into the spin current flowing up into the MTJ. The magnetic field is applied to the in-plane, 60° tilted from the long axis of the ellipse pillar. (b) In-plane magnetoresistance (MR) of the MTJ without spin current injection ($I_{Ta}$=0 mA, $I_{MTJ}=$ 20 µA). The magnetic field direction is the same as in Fig.1 (a). The black arrow shows the magnetic field at which the oscillation spectrum is measured. (c) Frequency dependence of $S_{11}$(reflection) of the transmission line alone measured without sample (the end of the amplifier is loaded and that of the bias-T is opened). Inset shows the schematic of the measurement setup for $S_{11}$.

**Figure 2:** Power spectra obtained for $I_{Ta}=$ (a) 0 mA, (b) 0.5 mA, (c) 1.0 mA, and (d) 1.5 mA. The dotted line in panel (d) is the envelope of the multipeak. The frequency spacing between each peak is approximately 105 MHz. (e): Image plot of the power spectrum as functions of the frequency and $I_{Ta}$.

**Figure 3:** $I_{Ta}$ dependence of (a) the total power of the STO main peak and (b) estimated FWHM. The black arrows at $I_{Ta}=0.5$ mA and 1.2 mA show the kink where the spectral feature shows a qualitative change.

**Figures**

(a)

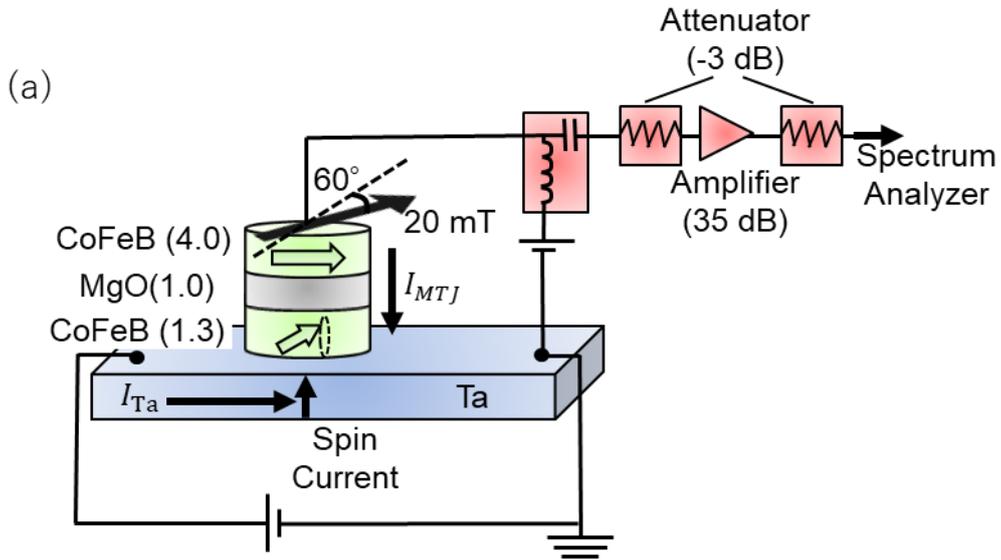

(b)

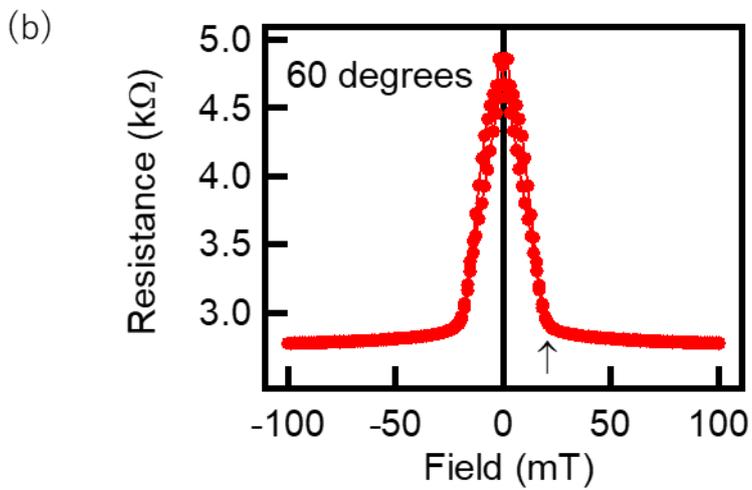

(c)

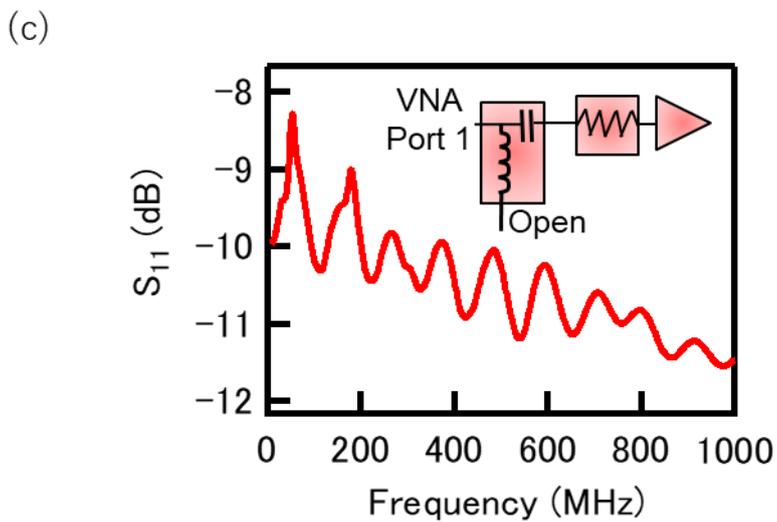

Figure 1 Iwakiri *et.al.,*

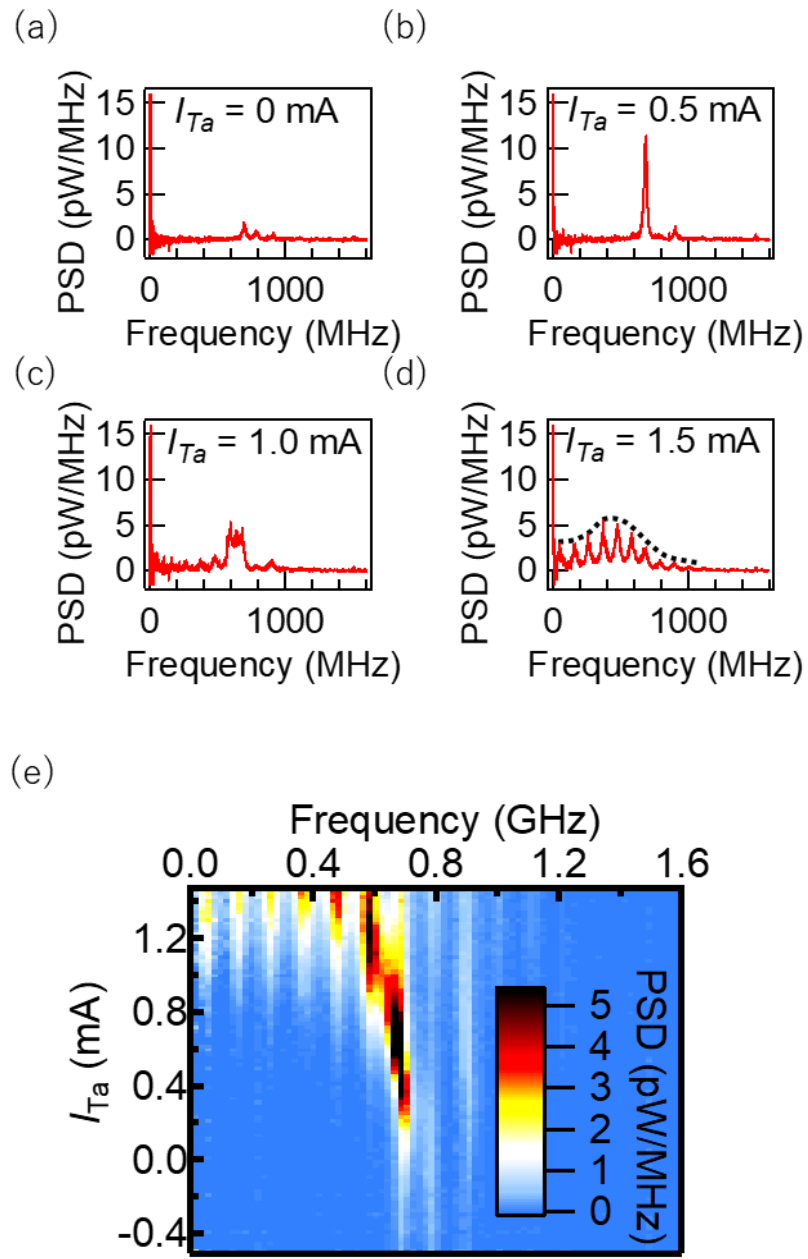

Figure 2 Iwakiri *et.al.*,

(a)

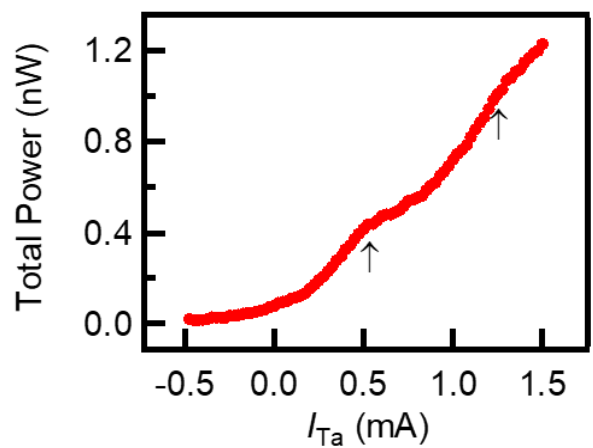

(b)

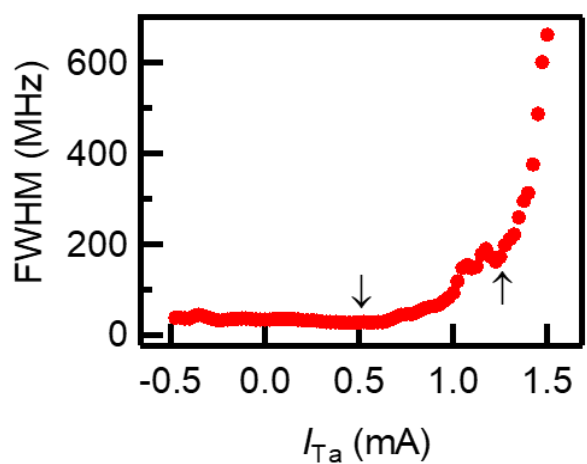

Figure 3 Iwakiri *et.al.*.

15